\documentclass[aps,twocolumn,groupedaddress,showpacs]{revtex4}
\usepackage{graphicx}
\newcommand\beq{\begin{eqnarray}}
\newcommand\eeq{\end{eqnarray}}
\newcommand\qom{\frac{q}{M}}
\newcommand\qove{\frac{q}{E_{\nu}}}
\newcommand\qovm{\frac{q}{2M}}
\newcommand\me{\frac{m_{\nu}}{E_{\nu}}}
\newcommand\fip{\frac{\mid \phi_{\mu}(0) \mid^{2}}{4\pi}}

\newcommand\scr{\mid C_{S}^{R} \mid^{2}}
\newcommand\vlsr{C_{V}^{L}C_{S}^{R*}}

\newcommand\pmo{{\bf |P_{\mu}|}}
\newcommand\pq{{\bf (\hat{P}_{\mu}\cdot \hat{q})}}
\newcommand\pqp{{\bf|P_{\mu}|(\hat{P}_{\mu}\cdot \hat{q})}}

\newcommand\pqkp{{\bf|P_{\mu}|(\hat{P}_{\mu}\cdot \hat{q})^{2}}}
\newcommand{\pqkpj}{{\bf|P_{\mu}|((\hat{P}_{\mu}\cdot \hat{q})^{2} } - 1)}
\begin{document}
\title{Right-handed Neutrinos in
Low-Energy  Neutrino-Electron Scattering}
\author{W. Sobk\'ow}
\email{sobkow@rose.ift.uni.wroc.pl}
\affiliation{Institute of Theoretical Physics, University of Wroc\l{}aw,
Pl. M. Born 9, PL-50-204~Wroc\l{}aw, Poland}
\date{\today}
\begin{abstract}
In this paper a scenario admitting the participation  of the
exotic   scalar  coupling of the right-handed neutrinos in
addition to the standard  vector and axial couplings of the
left-handed neutrinos in the weak interactions is considered. The
research is based on the low-energy $(\nu_{\mu} e^{-})$ and
$(\nu_{e} e^{-})$ scattering processes. The main goal is to show
how the presence of the right-handed neutrinos in the above
processes changes the laboratory differential cross section in
relation to the Standard Model prediction. Both processes are
studied at the level of the four-fermion point interaction.
Neutrinos are assumed to be polarized Dirac fermions and to be
massive. In the laboratory differential cross section, the new
interference term between the standard vector coupling of the
left-handed neutrinos and exotic scalar  coupling of the
right-handed neutrinos  appears which does not vanish in the
limit of vanishing neutrino mass. This additional contribution,
including information on the transverse components of neutrino
polarization, generates the azimuthal asymmetry in the angular
distribution of the recoil electrons. This regularity would be a
signature of the presence of the right-handed neutrinos in the
neutrino-electron scattering.
\end{abstract}
\maketitle
The $(V-A)$ structure  of weak interactions describes only
what has been measured so far. We mean here  the measurement of
the electron helicity \cite{Bob}, the indirect measurement of the
neutrino helicity \cite{Gold}, the asymmetry in the distribution
of the electrons from $\beta$-decay \cite{CWu} and the experiment
with muon decay \cite{Gar} which confirmed parity violation
\cite{Lee}.
Feynman, Gell-Mann and independently Sudarshan, Marshak
\cite{Gell} established that only left-handed vector $ V$, axial
$ A$ couplings can take part in weak interactions because this
yields the maximum symmetry breaking under space inversion, under
charge conjugation; the two-component neutrino theory of negative
helicity;  the conservation of the combined symmetry $ CP$ and of
the lepton number.
In consequence it led to the conclusion that produced neutrinos
in $V-A$ interaction can only be left-handed. However Wu
\cite{SWu} indicated  that  both  standard left-handed $(V,
A)_{L}$ couplings and exotic right-handed $(S, T, P)_{R}$
couplings may be responsible for the negative electron helicity
observed in $\beta$-decay. It would mean that generated neutrinos
in $(S, T, P)$ interactions  may also be {\it right-handed}
(antineutrinos {\it left-handed}, respectively). Recent tests do
not provide a unique
 answer as to the presence of  the exotic weak interactions.
\par So Shimizu {\it et al.}  \cite{Shimizu} determined the ratio of
the strengths of scalar and tensor couplings to the standard
vector coupling in $K^{+}\rightarrow \pi^{0}+e^{+}+\nu_{e}$ decay
at rest assuming the only left-handed neutrinos for all the
interactions. Their results indicated the compatibility with the
Standard Model (SM) \cite{Glashow,Wein,Salam} prediction. Bodek
{\it et al.}  at the PSI \cite{Bodek}  looked for the evidence of
the violation of time reversal invariance measuring $T$-odd
transverse components of the positron polarization in
$\mu^{+}$-decay. They also admitted the presence of the only
left-handed neutrinos produced in the standard $V-A$ and  scalar
interactions. The recent results presented by the DELPHI
Collaboration \cite{Delphi} concerning the measurement of the
Michel parameters and the neutrino helicity in $\tau$ lepton
decays indicated the consistency with the standard $V-A$ structure
of the charged current weak interaction. However on the other
hand, the achieved precision of measurements still admits the
deviation from the pure $V-A$ interaction, i.e. the possible
participation of the exotic couplings of the right-handed
neutrinos beyond the SM. There exist the models of the spontaneous
symmetry breaking under time reversal  in which the non-standard
scalar weak interaction can appear \cite{TDLee}. Recently
Berezhiani {\it et al.} \cite{Berez} analysed the scenario with
the  participation of the non-standard interactions of neutrinos
with electrons in the case of solar neutrinos.
\par It is necessary to carry out the new  high-precision tests of
the  Lorentz structure and of the handedness structure of the
weak interactions at low energies in which the {\it transverse
components of the neutrino polarization} would be measured,
because in the conventional observables  the interference terms
between the standard $(V, A)_{L}$  and   the exotic $(S, T,
P)_{R}$ couplings vanish in the limit of vanishing neutrino mass
\cite{Wolf,Miran,Sobkow}.
 Frauenfelder {\it et al.} \cite{spin} pointed out that one has to
measure either the neutrino polarization (spin) or the
neutrino-electron correlations to determine the full Lorentz
structure of the weak interactions. Because the direct measurement
of the transverse neutrino  polarization is difficult now,  the
low-energy neutrino-electron scattering as the detection process
of the exotic couplings  of the right-handed neutrinos can be
used. In practice, the low-energy strong and polarized neutrino
source (e. g. $^{51}Cr$ or other artificial neutrino sources)
could be used to search for the exotic effects. The process of the
neutrino-electron scattering can also be used to search for the
neutrino magnetic moments and to probe the flavor composition of
a (anti)neutrino beam \cite{Minkowski}. Barbieri and Fiorentini
\cite{Barbieri} analysed the conversions $\nu_{eL} \rightarrow
\nu_{eR}$ in the Sun as a result of spin-flip by a toroidal
magnetic field in the convective zone. They  calculated the
differential cross section for the $(\nu_{e} e^{-})$ scattering
with the solar $^{8}B$ neutrinos, assuming that the initial
neutrino flux is a mixture of both the left-handed neutrinos and
the right-handed ones. In this case  an interference term between
the weak interaction and electromagnetic amplitudes, proportional
to $\mu_{\nu}$, appears which generates the azimuthal asymmetry
in the recoil electron event rates.
 Pastor
{\it et al.} \cite{Pastor} calculated the azimuthal asymmetry for
the low-energy pp-neutrinos and discussed the sensitivity of
planned solar experiments to the expected azimuthal asymmetries in
event number. Beacom nad Vogel \cite{Beacom} derived  a new limit
on the neutrino magnetic moment using the 825-days SuperKamiokande
solar neutrino data: $|\mu_{\nu}|\leq 1.5 \cdot 10^{-10} \mu_{B}$
at $90 \% $ CL compatible with the existing reactor limit of
$|\mu_{\nu}|\leq 1.8 \cdot 10^{-10} \mu_{B}$.
\par The first concept of the use of the artificial neutrino
source comes from Alvarez who proposed a $^{65}Zn$ \cite{Alvarez}.
The $^{51}Cr$ and $^{37}Ar$ neutrino sources were proposed by
Raghavan \cite{Raghavan} in 1978 and Haxton \cite{Haxton} in 1988,
respectively. The idea of the use artificial neutrino source to
search for the neutrino magnetic moments was first proposed by
Vogel and Engel \cite{Vogel}. The strong $^{51}Cr$ source was used
for the calibration of the GALLEX neutrino experiment
\cite{Gallex}. Miranda {\it et al.} \cite{Miran} proposed the use
of the $^{51}Cr$ source to probe the gauge structure of the
electroweak interaction. Currently at Gran Sasso, the Borexino
neutrino experiment \cite{Miranda} with the unpolarized $^{51}Cr$
source is designed to search for the neutrino magnetic moment.
This experiment will use the $(\nu_{e} e-)$ scattering as the
detection process. There are also proposed the other experiments
to test the non-standard properties of neutrinos with the use of
both the solar neutrinos and neutrinos coming from the
beta-radioactive neutrino sources: the Hellaz \cite{Hellaz}, the
Heron \cite{Heron} and the new project with the use of the tritium
neutrino emitter \cite{Rashba,Trofimov}.
\par
The main  goal is to show how the presence of the right-handed
neutrinos in the neutrino-electron scattering changes the
laboratory differential cross section in relation to the Standard
Model prediction.
\par In
our considerations the system of natural units with $\hbar=c=1$,
Dirac-Pauli representation of the $\gamma$-matrices and the $(+,
-, -, -)$ metric are used \cite{Mandl}.
\par
\begin{figure}
\includegraphics[width=8.5cm,angle=0]{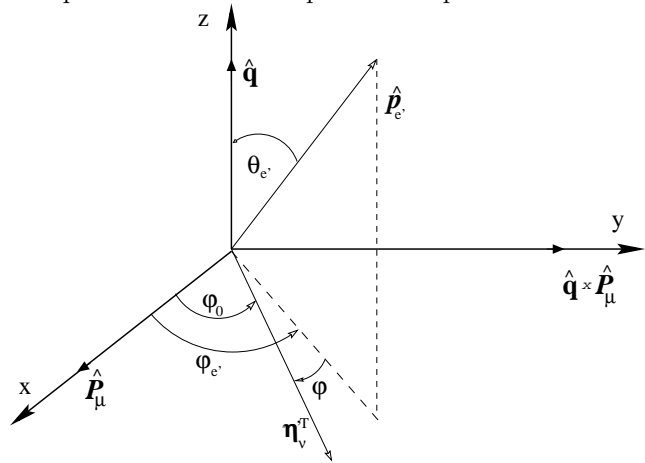}
\caption{Coordinate system conventions.}
 \label{wsp}
\end{figure}

 The research is based on the
low-energy $(\nu_{\mu} e^{-})$ and $(\nu_{e} e^{-})$ scattering
processes. Now, we will analyse  the $(\nu_{\mu}e^{-})$
scattering. This process is studied at the level of the
four-fermion point (contact) interaction. Muon-neutrinos are
assumed  to be massive Dirac fermions and to be polarized. In
these considerations, the incoming neutrinos come from the
muon-capture, where the production plane is spanned by the
direction of the initial muon polarization ${\bf {\hat P}_{\mu}}$
and of the outgoing neutrino momentum ${\bf \hat{q}}$
 (this is a production process). ${\bf {\hat P}_{\mu}}$
and ${\bf \hat{q}}$ are assumed to be perpendicular to each
other, Fig. \ref{wsp}. It is important from the experimental
point of view because one has the unique situation as to the
possible participation of the right-handed neutrinos.
The initial neutrino flux is the mixture of the left-handed
neutrinos produced in the standard $V-A$ charged weak interaction
and the right-handed ones produced in the exotic scalar $S$
charged weak interaction. Govaerts and Lucio-Martinez
\cite{Govaerts} considered the nuclear muon capture on the proton
and $^{3} He $ both within and beyond SM admitting the most
general Lorentz invariant four-fermion contact interaction. They
calculated the different observables assuming the Dirac massless
neutrino. In my paper, the minimal version of the extension of
the standard $V-A$ structure  is analysed to indicate the
possibility of new-type high-precision tests of the Lorentz
structure of the charged and neutral weak interactions.  The
amplitude for the $\mu^{-}$-capture and the formula on the
magnitude of the transverse neutrino polarization vector
$|\mbox{\boldmath $\eta_{\nu }^{T}$}|$, in the limit of vanishing
neutrino mass, are as follows:
\begin{widetext}
\beq {\cal H_{\mu^-}} & = & C_{V}^{L}({\overline\Psi}_{\nu}
    \gamma_{\lambda}(1 - \gamma_{5})\Psi_{\mu})
    ({\overline\Psi}_{n}\gamma^{\lambda}\Psi_{p}) \\
&  &  \mbox{} + C_{A}^{L}({\overline\Psi}_{\nu}
    i\gamma_{5}\gamma_{\lambda}(1 - \gamma_{5})\Psi_{\mu})
    ({\overline\Psi}_{n}i\gamma^{5}\gamma^{\lambda}\Psi_{p})
 +  C_{S}^{R}({\overline\Psi}_{\nu}(1 - \gamma_{5})\Psi_{\mu})
  ({\overline\Psi}_{n}\Psi_{p}), \nonumber\\
|\mbox{\boldmath $\eta_{\nu }^{T}$}| & = & \frac{\sqrt{{\bf <
S_{\nu}\cdot({\hat P}_{\mu}\times {\hat q}) >}_{f}^{2} + {\bf <
S_{\nu}\cdot {\hat P}_{\mu}>}_{f}^{2}}}{s <\bf 1>_{f}}
 =  \pmo
|\frac{C_{S}^{R}}{C_{V}^{L}}|(1+\qovm) \\
&  & \times  \{s[(3+\qom)|\frac{C_{A}^{L}}{C_{V}^{L}}|^{2} +
(1+\qom) + |\frac{C_{S}^{R}}{C_{V}^{L}}|^{2}
- 2\qom|\frac{C_{A}^{L}}{C_{V}^{L}}| cos(\alpha_{AV}^{L})]\}^{-1}, \nonumber\\
{\bf <S_{\nu}\cdot({\hat P}_{\mu}\times\hat{q})>}_{f} & \equiv &
Tr\{{\bf S_{\nu}\cdot({\hat P}_{\mu}\times\hat{q})}\rho_{f}\}
 =  - \fip\pmo (1 + \qovm)Im(\vlsr), \\
{\bf <S_{\nu}\cdot{\hat P}_{\mu}>}_{f} & \equiv & Tr\{ {\bf
S_{\nu}\cdot{\hat P}_{\mu}}\rho_{f}\}
   =   \fip\pmo (1 + \qovm)Re(\vlsr),
\eeq
\end{widetext}
 where $C_{V}^{L}, C_{A}^{L}, C_{S}^{R}$ - the complex fundamental coupling constants for the
standard  vector $V$, axial $A$ and exotic scalar $S$ weak
interactions denoted respectively to the outgoing neutrino
handedness; s - the neutrino spin (s=1/2);
 ${\bf <S_{\nu}\cdot({\hat P}_{\mu}\times\hat{q})>}_{f},
{\bf <S_{\nu}\cdot{\hat P}_{\mu}>}_{f}, {\bf <1>_{f}}$ - the
$T-$odd and $T-$even transverse components of neutrino
polarization and the probability of muon capture, respectively;
${\bf S_{\nu}}$ - the operator of the neutrino spin; $\rho_{f}$ -
the density matrix of the final state;
$q, M$ - the value of the neutrino momentum and the nucleon mass;
$\pmo$ - the value of the muon polarization in $1s$ state;
$\phi_{\mu}(0)$ - the value of the large radial component of the
muon Dirac bispinor for $r=0$; $\alpha_{AV}^{L} \equiv
\alpha_{A}^{L} - \alpha_{V}^{L} $ - the relative phase between the
standard $C_{A}^{L}$ and  $C_{V}^{L}$ couplings. The above
neutrino observables were calculated with the use of the density
matrix of the final state.
\par It can be seen that
the neutrino  observables consist exclusively of the interference
term  between the standard  $C_{V}^{L}$ coupling
 and exotic  $C_{S}^{R}$ one.
 It can be understood as the interference between the neutrino
waves of negative and positive handedness. There is no
 contribution to these observables from the SM in which neutrinos
 are only left-handed and  massless. The induced couplings
generated by the dressing of hadrons are neglected as their
presence does not change qualitatively the conclusions about
transverse neutrino polarization. These couplings enter additively
to the fundamental $C_{V,A}^{L}$ couplings (see Appendix A). The
mass terms in the above neutrino observables give a very small
contribution in relation to the main one coming from the
interference term (see Appendix A) and they can be neglected in
the considerations. General results for the neutrino observables
in the limit of vanishing neutrino mass, when ${\bf {\hat
P}_{\mu}}$ and ${\bf \hat{q}}$ are not perpendicular to each
other, are presented in Appendix B.  If one assumes the
production of the only left-handed neutrinos in all the
interactions, i.e. both for the standard $V-A$  and scalar $S$
interactions, there is no interference between the $C_{V,A}^{L}$
and $C_{S}^{L}$ couplings in the
 limit of vanishing neutrino mass. In this case $\mbox{\boldmath
$\eta_{\nu}^{T}$}=0$. The situation is qualitatively different in
the case of the longitudinal and transverse neutron polarization
and of the longitudinal neutrino polarization, where all the
interference terms between the $C_{V, A}^{L}$
 and   $C_{S}^{R}$ couplings are suppressed by the neutrino mass (see
Appendix C, D) and the standard $C_{V, A}^{L}$ couplings of the
left-handed neutrinos  dominate in agreement  with the SM.
Therefore, the neutrino observables in which such difficulties do
not appear are proposed. In this way, the conclusions as to the
existence of the right-handed neutrinos can depend on  the type of
measured observables. \par  Our coupling constants $C_{V,A}^{L},
C_{S}^{R}$ can be expressed by Fetscher's couplings
$g^{\gamma}_{\epsilon \mu}$ for the normal and inverse muon decay
\cite{Data}, assuming the universality of weak interactions.
Here, $\gamma= S, V, T$ indicates a scalar, vector, tensor
interaction; $\epsilon, \mu=L, R$ indicate the chirality of the
electron or muon and the neutrino chiralities are uniquely
determined for given $\gamma, \epsilon, \mu$. We get the
following relations: \beq C_{V}^{L} &=& A(g_{LL}^{V} +
g_{RL}^{V}), \\ -C_{A}^{L} &=& A(g_{LL}^{V} - g_{RL}^{V}),
\nonumber\\ C_{S}^{R} &=& A(g_{LL}^{S} + g_{RL}^{S}), \nonumber
\eeq where $A\equiv(4G_{F}/\sqrt{2})cos\theta_{c}$, $G_{F}=
1.16639(1)\times 10^{-5}GeV^{-2}$  is the Fermi coupling constant
\cite{Data}, $\theta_{c}$ is the Cabbibo angle.
In this way,  the lower  limits on the $C^{L}_{V,A}$ and upper
limit on the $C_{S}^{R}$  can be calculated, using the current
data \cite{Data}; $|C_{V}^{L}|>0.850 \,A, \;|C_{A}^{L}|>1.070 \,A,
\;|C_{S}^{R}|<0.974 \,A$. In consequence, one gives the upper
bound on the  magnitude  of the transverse neutrino polarization
vector proportional
 to the value of the muon polarization;
 $|\mbox{\boldmath $\eta_{\nu }^{T}$}|\leq 0.318 \pmo $ (for
$\alpha_{AV}^{L}=\pi$).
 The obtained limit has to be divided by the $\pmo$
 to have the upper bound
on the physical  value of the transverse  neutrino polarization
vector generated by the exotic scalar interaction;
$|\mbox{\boldmath $\eta_{\nu}^{' T}$}| = |\mbox{\boldmath
$\eta_{\nu }^{T}$}|/ \pmo \leq 0.318$.
\par From the above, we see that the analysis for the
$\mu^{-}$-capture \cite{Sobkow} led to the conclusion that the
production of the right-handed neutrinos in the exotic scalar
interaction manifests the non-vanishing  transverse components of
the neutrino polarization in the limit of vanishing neutrino mass,
so one expects the similar regularity in the $(\nu_{\mu}e^{-})$
scattering.
 We assume that the incoming left-handed neutrinos are detected
in the $V-A$   weak interaction, while the initial right-handed
neutrinos are detected in the  exotic scalar $S$  weak
interaction. In  the final state all the neutrinos are
left-handed.  One assumes that the initial neutrino beam has the
assigned direction of the transverse neutrino polarization with
the respect to the production plane, Fig.\ref{wsp}.  The couplings
constants are denoted as $g_{V}^{L}, g_{A}^{L}$ and $g_{S}^{R}$
respectively to the incoming neutrino handedness: \beq { \cal M}
&=&
\frac{G_{F}}{\sqrt{2}}\{(\overline{u}_{e'}\gamma^{\alpha}(g_{V}^{L}
- g_{A}^{L}\gamma_{5})u_{e}) (\overline{u}_{\nu_{\mu'}}
\gamma_{\alpha}(1 - \gamma_{5})u_{\nu_{\mu}})\nonumber \\ &  &
\mbox{} +
\frac{1}{2}g_{S}^{R}(\overline{u}_{e'}u_{e})(\overline{u}_{\nu_{\mu'}}
(1 + \gamma_{5})u_{\nu_{\mu}})\},
 \eeq
 where $ u_{e}$ and  $u_{e'}$
$(u_{\nu_{\mu}}\;$ and $\; u_{\nu_{\mu'}})$ are the Dirac
bispinors of the initial and final electron (neutrino)
respectively. \par The analysis of the general Lorentz invariant
four-fermion point interaction for the $2\rightarrow 2$ processes
involving two neutrinos and two charged fermions is presented by
\cite{Mendy}.
\par To describe $(\nu_{\mu}e^{-})$ scattering the following
observables are used: $\mbox{\boldmath $\eta_{\nu}$}$ - the full
3-vector of the initial neutrino polarization in the rest frame,
${\bf q}$ - the incoming neutrino momentum, ${\bf p_{e'}}$ - the
outgoing electron momentum.
\par The laboratory differential cross
section for the $\nu_{\mu}e^{-}$ scattering, in the limit of
vanishing neutrino mass, is of the form:
\begin{widetext}
\beq \label{przekr} \lefteqn{\frac{d^{2} \sigma}{d y d \phi_{e'}}
= (\frac{d^{2} \sigma}{d y d \phi_{e'}})_{(V, A) } + (\frac{d^{2}
\sigma}{d y d \phi_{e'}})_{(S)}
+ (\frac{d^{2} \sigma}{d y d \phi_{e'}})_{(V S)},}\\
(\frac{d^{2} \sigma}{d y d \phi_{e'}})_{(V, A)} &=& B \{
(1-\mbox{\boldmath $\eta_{\nu}$}\cdot\hat{\bf q} )[(g_{V}^{L} +
g_{A}^{L})^{2}
 + (g_{V}^{L}- g_{A}^{L})^{2}(1-y)^{2}
- \frac{m_{e}y}{E_{\nu}}((g_{V}^{L})^{2} - (g_{A}^{L})^{2})]\}, \\
(\frac{d^{2} \sigma}{d y d \phi_{e'}})_{(S)} &=& \mbox{}
B\{\frac{1}{8}y(y+2\frac{m_{e}}{E_{\nu}})
[ |g_{S}^{R}|^{2}(1+\mbox{\boldmath $\eta_{\nu}$}\cdot\hat{\bf q})]\}, \\
(\frac{d^{2} \sigma}{d y d \phi_{e'}})_{(V S)} &=& \mbox{}
B\{\sqrt{y(y+2\frac{m_{e}}{E_{\nu}})}[-\mbox{\boldmath
$\eta_{\nu}$}\cdot({\bf \hat{p}_{e'} \times
\hat{q}})Im(g_{V}^{L}g_{S}^{R*}) + (\mbox{\boldmath
$\eta_{\nu}$}\cdot
{\bf \hat{p}_{e'}}) Re(g_{V}^{L}g_{S}^{R*})] \\
&& \mbox{} - y(1+\frac{m_{e}}{E_{\nu}}) (\mbox{\boldmath
$\eta_{\nu}$}\cdot\hat{\bf q}) Re(g_{V}^{L}g_{S}^{R*})\},\nonumber
\eeq
\end{widetext}
\beq B & \equiv & \frac{E_{\nu}m_{e}}{(2\pi)^{2}}
\frac{G_{F}^{2}}{2}, \\ y & \equiv &
\frac{E_{e'}^{R}}{E_{\nu}}=\frac{m_{e}}{E_{\nu}}\frac{2cos^{2}\theta_{e'}}
{(1+\frac{m_{e}}{E_{\nu}})^{2}-cos^{2}\theta_{e'}}, \eeq where
$y$- the ratio of the kinetic energy of the recoil electron
$E^{R}_{e'}$  to the incoming neutrino energy $E_{\nu}$,
$\theta_{e'}$ - the angle between the direction of the outgoing
electron momentum  $ \hat{\bf p}_{e'}$  and the direction  of the
incoming neutrino momentum $\hat{\bf q}$ (recoil electron
scattering angle), $m_{e}$ - the electron mass, $ \mbox{\boldmath
$\eta_{\nu}$}\cdot\hat{\bf q}$ - the longitudinal polarization of
the incoming neutrino, $\phi_{e'}$ - the angle between the
production plane and the reaction plane spanned by the $ \hat{\bf
p}_{e'}$ and $ \hat{\bf q}$ (see Fig. \ref{wsp}). All the
calculations are made with the Michel-Wightman density matrix
\cite{Michel} for the polarized incoming neutrinos in the limit of
vanishing neutrino mass (see Appendix E).
\par
\begin{figure}
\includegraphics[width=8.5cm,angle=0]{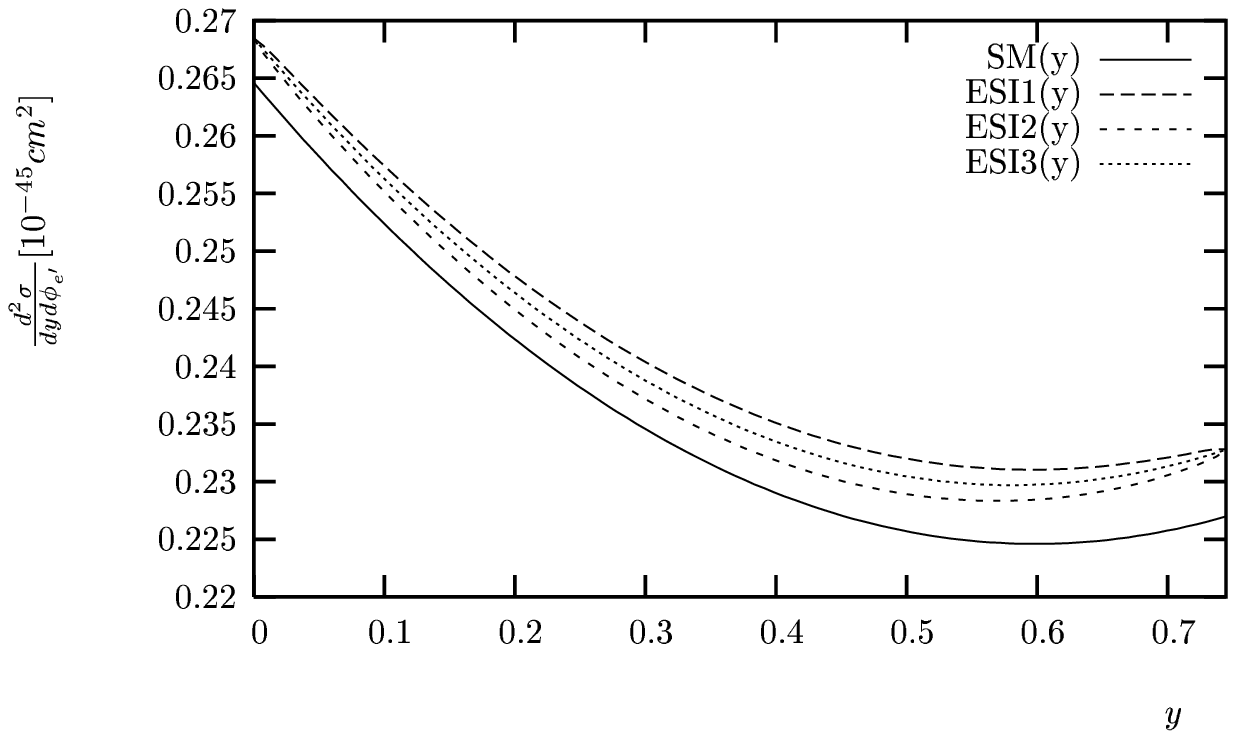}
\caption{Plot of the $\frac{d^{2} \sigma}{d y d \phi_{e'}}$ as a
function of $y$ for the $(\nu_{\mu}e^{-})$ scattering; a) SM
with the left-handed neutrino (SM(y) - solid line), b) the case
of the exotic scalar coupling  of the right-handed neutrino for
$\phi-\alpha=0$ (ESI1(y)- dashed line), for $\phi-\alpha=\pi$
(ESI2(y) - dashed line) and for $\phi-\alpha=\pi/2$ (ESI3(y) -
dotted line), respectively.}
 \label{wykr1}
\end{figure}

 It can be
noticed that  the main non-standard contributions to the
laboratory differential cross section come from the interference
between the standard left-handed vector $g_{V}^{L}$ coupling and
exotic right-handed scalar $g_{S}^{R}$ coupling, whose  occurrence
does not depend explicitly on the neutrino mass. This interference
may be understood as the interference between the neutrino waves
of the same handedness for the  final neutrinos.
The similar regularity as for $\mu^{-}$-capture  appeared here
\cite{Sobkow}.
The correlation $\mbox{\boldmath $\eta_{\nu}$}\cdot {\bf
\hat{p}_{e'}}$ proportional to $Re(g_{V}^{L}g_{S}^{R*})$ lies in
the  reaction plane and it  includes both  $T$-even longitudinal
and transverse components of the initial neutrino polarization:
\beq (\mbox{\boldmath $\eta_{\nu}$}\cdot {\bf \hat{p}_{e'}})&=&
cos\theta_{e'}(\mbox{\boldmath $\eta_{\nu}$}\cdot\hat{\bf q}) +
(\mbox{\boldmath $\eta_{\nu }^{' T}$}\cdot{\bf {p}_{e'}^{T}}),
 \eeq
where index ''T'' describes  transverse components of the
outgoing electron momentum  and of the incoming neutrino
polarization, respectively.
The other correlation  $\mbox{\boldmath $\eta_{\nu}$}\cdot({\bf
\hat{p}_{e'}\times \hat{q}})$ proportional to
$Im(g_{V}^{L}g_{S}^{R*})$ lies along the direction perpendicular
to the reaction plane and it includes only $T$-odd transverse
component of the initial neutrino polarization: \beq
\mbox{\boldmath $\eta_{\nu}$}\cdot({\bf \hat{p}_{e'} \times
\hat{q}})&=& \mbox{\boldmath $\eta_{\nu}^{' T}$}\cdot({\bf
\hat{p}_{e'} \times \hat{q}}). \eeq It can be shown that in the
full interference term, Eq. (10), the contributions from the
longitudinal components of the neutrino polarization annihilate,
and in consequence one gives the interference  including only the
transverse components of the initial neutrino polarization, both
$T$-even and $T$-odd: \beq \label{inter} (\frac{d^{2} \sigma}{d y
d \phi_{e'}})_{(VS)}  &=&
B\{\sqrt{\frac{m_{e}}{E_{\nu}}y[2-(2+\frac{m_{e}}{E_{\nu}})y]}\\
&& \times |g_{V}^{L}||g_{S}^{R}||\mbox{\boldmath $\eta_{\nu}^{'
T}$}|cos(\phi-\alpha)\}, \nonumber \eeq where $\alpha
\equiv\alpha_{V}^{L} - \alpha_{S}^{R} $ - the relative phase
between the  $g_{V}^{L}$ and  $g_{S}^{R}$ couplings, $\phi$ - the
angle between the reaction plane and the transverse neutrino
polarization vector and it is connected with the $\phi_{e'}$ in
the following way; $\phi=\phi_{0}-\phi_{e'}$, where $\phi_{0}$ -
the angle between the production plane and the transverse neutrino
polarization vector (see Fig. \ref{wsp}). It can be noticed that
the contribution from the interference between the $g_{V}^{L}$ and
$g_{S}^{R}$ couplings, involving the transverse neutrino
polarization components, will be substantial at lower neutrino
energies $E_{\nu}\leq m_{e}$  but negligibly small at large
energies and vanishes for $\theta_{e'}=0$ or $\theta_{e'}=\pi/2$.
The occurrence of the interference term in the cross section
depends on the relative phase between the angle $\phi$ and phase
$\alpha$ and does not vanish for $\phi - \alpha \not=\pi/2$. The
azimuthal asymmetry  in the angular distribution of the recoil
electrons is illustrated  in the Fig. \ref{wykr1}, when
$\phi-\alpha=0$ (ESI1(y) - dashed line), $\phi-\alpha=\pi$
(ESI2(y) - dashed line) and $\phi-\alpha=\pi/2$ (ESI3(y) - dotted
line), respectively.
\begin{figure}
\includegraphics[width=8.5cm,angle=0]{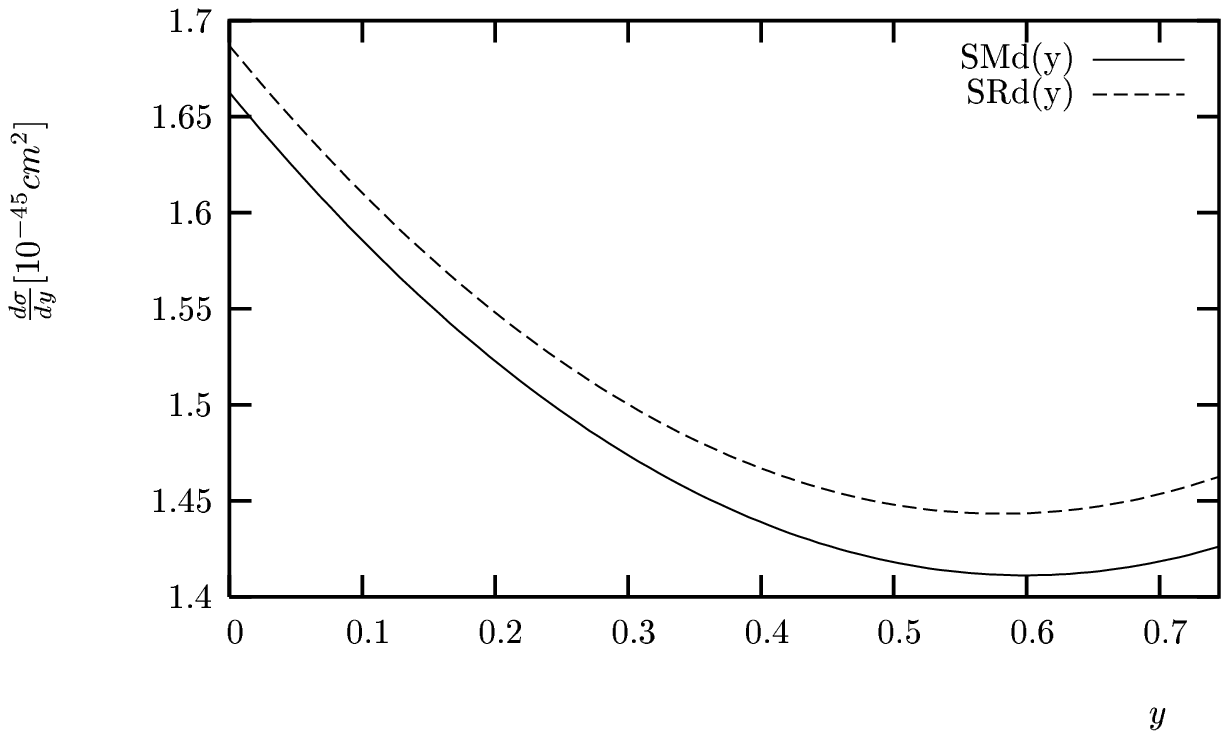}
\caption{Plot of the $\frac{d \sigma}{d y}$ as a function of $y$
for the $(\nu_{\mu}e^{-})$ scattering; a) SM  with the left-handed
neutrino (SMd(y) - solid line), b) the case of the exotic scalar
coupling of the right-handed neutrino  after integration over the
$\phi_{e'}$(SRd(y) - dashed line).}
 \label{wykr2}
\end{figure}
 All the plots (Fig.2 - Fig.5) both for the  $(\nu_{\mu}
e^{-})$ and $(\nu_{e} e^{-})$ scattering processes  are made for
$m_{e}= 511 \, keV, E_{\nu}=746 \, keV$ and $y \in [0, 0.745]$.
Because the right-handed neutrinos are produced and detected in
the exotic scalar $S$ interaction, one uses the same upper limit
on the $g_{S}^{R}$ as for the $C_{S}^{R}$, i. e.
$|g_{S}^{R}|<0.974 $, assuming the universality of weak
interaction. We take the upper bound on the $|\mbox{\boldmath
$\eta_{\nu}^{' T}$}|\leq 0.318$ for the neutrinos coming from the
muon-capture (however the phase $\alpha$ is still unknown). The
value $|\mbox{\boldmath $\eta_{\nu}^{' T}$}|=0.318$ is used to
get the upper limit on the expected effect from the right-handed
neutrinos in the cross section for the $(\nu_{\mu}e^{-})$
scattering. It means that the value of the longitudinal neutrino
polarization is equal to $\mbox{\boldmath $\eta_{\nu}^{l}$}\equiv
\mbox{\boldmath $\eta_{\nu}$}\cdot\hat{\bf q}=-0.948$.
The plot for the SM is made with the use of the present
experimental values for $g_{V}^{L}=-0.040\pm 0.015$,
$g_{A}^{L}=-0.507\pm 0.014$ \cite{Data}, when $\mbox{\boldmath
$\eta_{\nu}$}\cdot\hat{\bf q}= -1$, Fig. \ref{wykr1} and Fig.
\ref{wykr2} (SM(y), SMd(y) - solid lines). If one integrates over
the $\phi_{e'}$, the interference term vanishes and the cross
section consists of only two terms:
\beq \label{wyk2} \frac{d \sigma}{d y} & = & (\frac{d \sigma}{d
y})_{(V, A)} + (\frac{d \sigma}{d y})_{(S)},\\ (\frac{d \sigma}{d
y})_{(V, A)} &=& B' \{ (1-\mbox{\boldmath
$\eta_{\nu}$}\cdot\hat{\bf q})[(g_{V}^{L} + g_{A}^{L})^{2} \\ & +
& (g_{V}^{L}- g_{A}^{L})^{2}(1-y)^{2} \nonumber\\
 & - & \frac{m_{e}y}{E_{\nu}}((g_{V}^{L})^{2} - (g_{A}^{L})^{2})]\}, \nonumber \\
(\frac{d \sigma}{d y})_{(S)} &=& \mbox{} B'
\{\frac{1}{8}y(y+2\frac{m_{e}}{E_{\nu}}) \\ & \times &
[|g_{S}^{R}|^{2}(1+\mbox{\boldmath $\eta_{\nu}$}\cdot\hat{\bf
q})]\}, \nonumber
 \eeq
where $B' = 2\pi B$. The situation is illustrated in the Fig.
\ref{wykr2} (SRd(y) - dashed line).
 If the only left-handed neutrinos are produced in  the standard
$V-A$ interaction and non-standard scalar $S$ one, they should be
detected in the same interactions. In this case there is no
interference between the $(V, A)_{L}$ and $S_{L}$ couplings in
the differential cross section, when $m_{\nu}\rightarrow 0$, and
the angular distribution of the recoil electrons has the azimuthal
symmetry. Because the left-handed scalar $S_{L}$ coupling is
absent in the production process, so this scenario is not
considered for the $(\nu_{\mu}e^{-})$ scattering.
\begin{figure}
\includegraphics[width=8.5cm,angle=0]{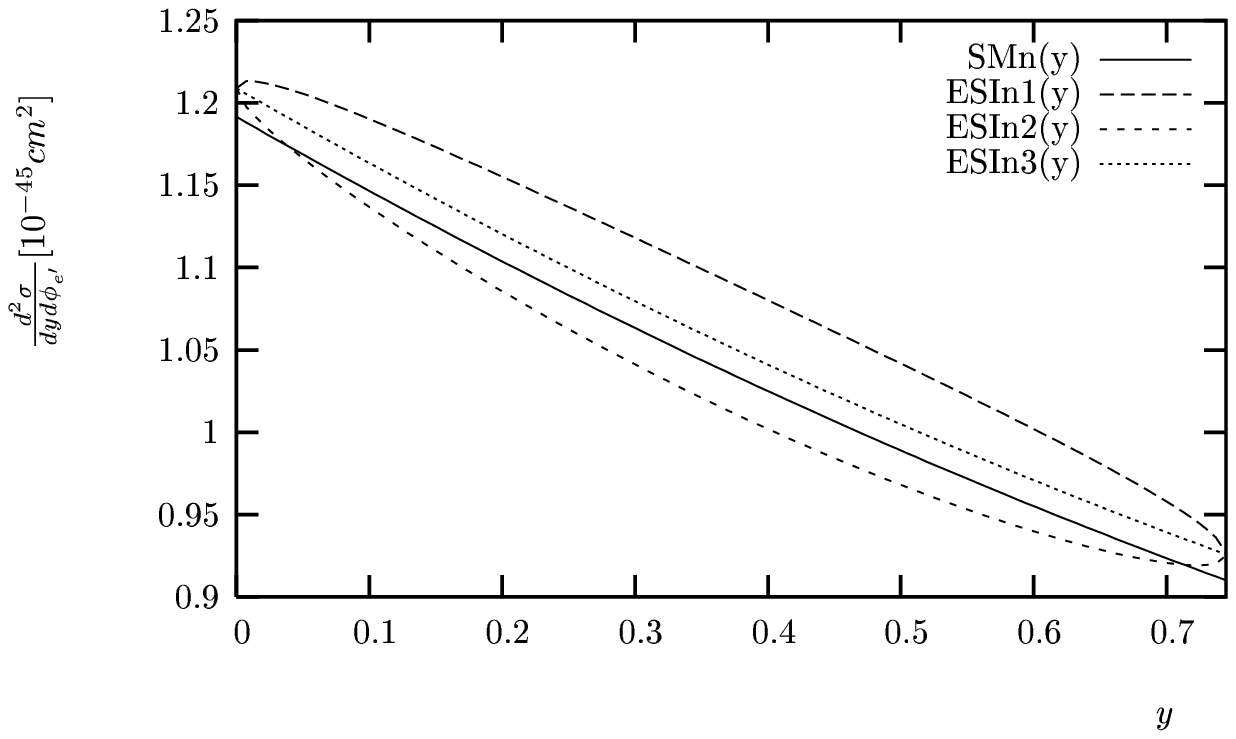}
\caption{Plot of the $\frac{d^{2} \sigma}{d y d \phi_{e'}}$ as a
function of $y$ for the $(\nu_{e} e^{-})$ scattering; a) SM  with
the left-handed neutrino (SMn(y) - solid line), b) the case of the
exotic scalar coupling of the right-handed neutrino for
$\phi-\alpha=0$ (ESIn1(y)- dashed line), for $\phi-\alpha=\pi$
(ESIn2(y) - dashed line) and for $\phi-\alpha=\pi/2$ (ESIn3(y) -
dotted line), respectively.}
 \label{wykr3}
\end{figure}
\begin{figure}
\includegraphics[width=8.5cm,angle=0]{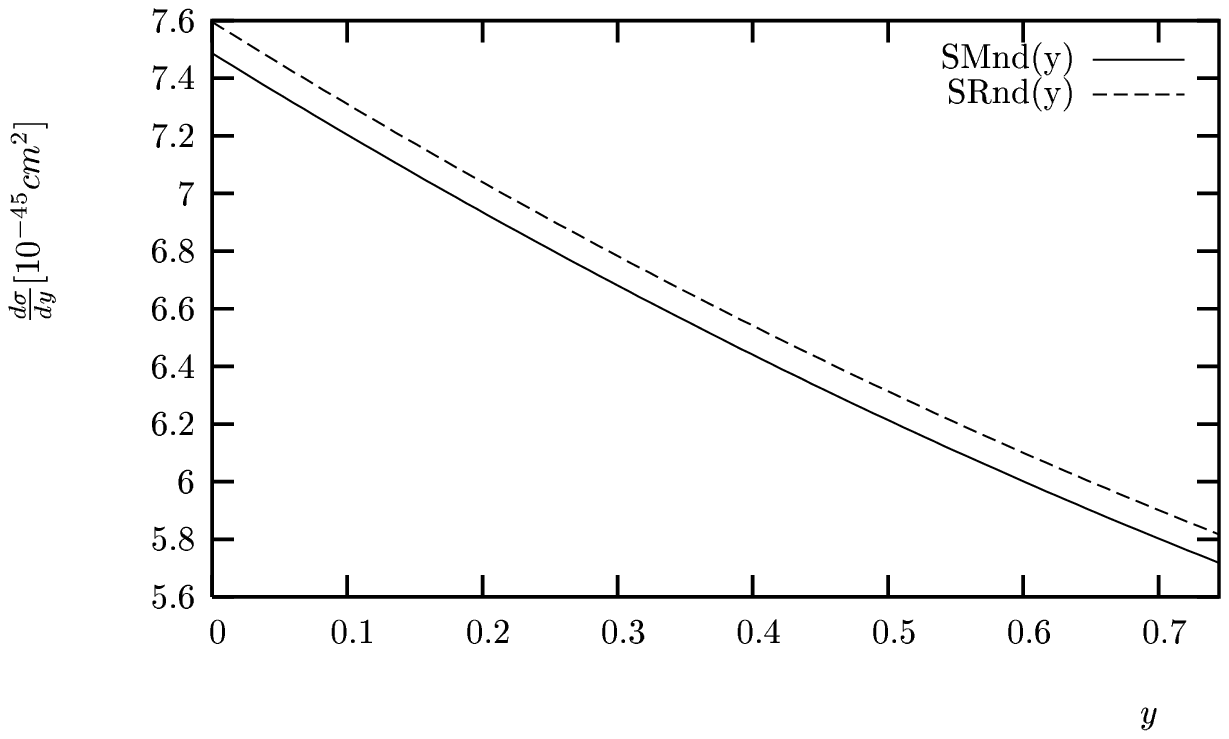}
\caption{Plot of the $\frac{d \sigma}{d y}$ as a function of $y$
for the $(\nu_{e}e^{-})$ scattering; a) SM  with the left-handed
neutrino (SMnd(y) - solid line), b) the case of the exotic scalar
coupling of the right-handed neutrino  after integration over the
$\phi_{e'}$ (SRnd(y) - dashed line).}
 \label{wykr4}
\end{figure}
\par Taking into account the possibilities of the
future low-energy neutrino experiments,  we consider the $(\nu_{e}
e^{-})$ process. If the azimuthal asymmetry in the cross section
for the $(\nu_{\mu} e^{-})$ scattering  appears, the similar
regularity for the $(\nu_{e} e^{-})$ process should occur, when
the electron-neutrinos come from  the polarized artificial
neutrino source ($^{51}Cr$). The $^{51}Cr$-decay proceeds by the
electron capture and is similar to the $\mu^{-}-$capture. It is
well-known that the $^{51}Cr$ decays with a $Q-$value of $751 \;
keV$ to the ground state of $^{51}V$ ($90.14 \%$ branching ratio)
and to its first excited state ($9.86 \pm 0.05\%$), which
deexcites to the ground state with the emission of a $320 \; keV$
$\gamma-$ray. The neutrino spectrum consists of four
monoenergetic lines: $746 \; keV \; (81 \%)$, $751 \; keV \; (9
\%)$, $426 \; keV \; (9 \%)$, $431 \; keV \; (1 \%)$ \cite{PTB}.
If the chromium neutrino source would be polarized, one would have
the fixed direction of the transverse neutrino polarization with
respect to the production plane spanned by the initial
polarization ${\bf J_{i}}$ of the $^{51}Cr$ and the outgoing
electron-neutrino momentum ${\bf q}$. The reaction plane is the
same as for the $(\nu_{\mu} e^{-})$ scattering. It would allow to
measure the azimuthal asymmetry in the angular distribution of
the recoil electrons. In the amplitude for the $(\nu_{e} e^{-})$
scattering, the couplings constants are denoted as $c_{V}^{L},
c_{A}^{L}$ and $c_{S}^{R}$ respectively to the incoming neutrino
handedness, where $c_{V}^{L}= g_{V}^{L}+1, \; c_{A}^{L}=
g_{A}^{L}+1$ (the charged current weak interaction is included).
The plot for the SM  is made with the same values of the standard
coupling constants as for the $(\nu_{\mu} e^{-})$ process, i. e.
$ c_{V}^{L}=-0.040 + 1$, $c_{A}^{L}=-0.507 + 1$, when
$\mbox{\boldmath $\eta_{\nu}$}\cdot\hat{\bf q}= -1$, Fig.
\ref{wykr3} and Fig. \ref{wykr4} (SMn(y), SMnd(y) - solid lines).
The upper limit on the $c_{S}^{R}$ is the same as for the
$g_{S}^{R}$, i. e. $|c_{S}^{R}|<0.974 $, assuming the
universality of weak interaction. We also use both
$|\mbox{\boldmath $\eta_{\nu}^{' T}$}|=0.318$ and $\mbox{\boldmath
$\eta_{\nu}^{l}$}\equiv \mbox{\boldmath $\eta_{\nu}$}\cdot\hat{\bf
q}=-0.948$ to obtain the upper bound on the possible effect from
the R-handed neutrinos for the $(\nu_{e} e^{-})$ scattering. The
azimuthal asymmetry  in the angular distribution of the recoil
electrons is illustrated in the Fig. \ref{wykr3} (ESIn1(y),
ESIn2(y) - dashed lines, ESIn3(y) - dotted line). If one
integrates over the $\phi_{e'}$, the interference term vanishes
and the cross section has the azimuthal symmetry, Fig.
\ref{wykr4} (SRnd(y) - dashed line).
\par In summary, it is  known that in the SM the angular distribution of the
recoil electrons does not depend on the $\phi_{e'}$. It is
necessary to observe the direction of the recoil electrons and to
analyse  all the possible reaction planes corresponding to the
given recoil electron scattering angle to verify if the azimuthal
asymmetry in the cross section appears. The regularity of this
type  would be a signature indicating the possible participation
of the right-handed neutrinos in the neutrino-electron scattering.
The future low-energy high-precision neutrino-electron scattering
experiments using the intense and polarized  neutrino source (e.
g. $^{51}Cr$ or other artificial neutrino  sources) would allow to
search for new effects coming from the {\sl right-handed
neutrinos} (to be published).
 \begin{widetext}
\appendix
\section{Muon neutrino observables}
The formulas for  the transverse components of neutrino
 polarization (T-odd and T-even, respectively) in case of  non-vanishing neutrino mass
 $(m_{\nu}\not = 0)$,
 when the induced couplings are enclosed and ${\bf
{\hat P}_{\mu}}$,  ${\bf \hat{q}}$ are  perpendicular to each
other, are as follows: \beq {\bf <S_{\nu}\cdot({\hat
P}_{\mu}\times\hat{q})>}_{f} & \equiv & Tr\{{\bf
S_{\nu}\cdot({\hat P}_{\mu}\times\hat{q})}\rho_{f}\}
 =  - \fip\pmo (\qove + \qovm)Im((C_{V}^{L} + 2 M g_{V}^{L})C_{S}^{R*}), \\
{\bf <S_{\nu}\cdot{\hat P}_{\mu}>}_{f} & \equiv & Tr\{ {\bf
S_{\nu}\cdot{\hat P}_{\mu}}\rho_{f}\}
   =   \fip\pmo \{(1 + \qove \qovm)Re((C_{V}^{L} + 2 M g_{V}^{L})C_{S}^{R*})\\
  & & \mbox{} + \frac{1}{2} \me( |C_{V}^{L} + 2 M g_{V}^{L}|^{2} - |C_{A}^{L} +
  m_{\mu}\qovm  g_{A}^{L}|^{2} + \scr)\}, \nonumber
\eeq where $g_{V}^{L}, g_{A}^{L}$ - the induced couplings of the
left-handed neutrinos, i.e. the weak magnetism and induced
pseudoscalar, respectively;
 $m_{\mu}, q, E_{\nu}, m_{\nu}, M$ - the muon mass, the value of the neutrino momentum, its
 energy,
 its mass and the nucleon mass. If $m_{\nu} \rightarrow 0, \; q/E_{\nu} \rightarrow 1 $  and
 the mass terms vanish in all the observables.
\section{General form of muon neutrino observables}
General results for the transverse components of the neutrino
polarization, in the limit of vanishing neutrino mass $(m_{\nu}
\rightarrow  0)$, when the induced couplings are enclosed and
${\bf {\hat P}_{\mu}}$, ${\bf \hat{q}}$ are not perpendicular to
each other, are as follows: \beq {\bf <
S_{\nu}\cdot(\hat{P}_{\mu}\times \hat{q})
>}_{f} & = &
 \mbox{} \fip \pqkpj(1 + \qovm) Im((C_{V}^{L} + 2 M
 g_{V}^{L})C_{S}^{R*}),\\
{\bf <S_{\nu}\cdot \hat{ P}_{\mu}>}_{f} & =  & \mbox{} \fip \{-
\pqkpj(1 + \qovm) Re((C_{V}^{L} + 2 M
g_{V}^{L})C_{S}^{R*}) \\
&& \mbox{} + \pqkp  [\frac{1}{2} \scr +  \qom Re((C_{V}^{L} +
2 M g_{V}^{L})(C_{A}^{L*} +  m_{\mu}\qovm g_{A}^{L*})) \nonumber \\
  & & \mbox{} + \frac{1}{2}(1 +  \qom) |C_{V}^{L} + 2 M g_{V}^{L}|^{2}
+ \frac{1}{2}(  \qom - 1) |C_{A}^{L} +
  m_{\mu}\qovm  g_{A}^{L}|^{2}] \nonumber \\
  & & \mbox{} + \pq[\frac{1}{2} \scr
+ \qom Re((C_{V}^{L} + 2 M g_{V}^{L})(C_{A}^{L*} +
  m_{\mu}\qovm g_{A}^{L*})) \nonumber \\
&& \mbox{} - \frac{1}{2} (1 +  \qom) |C_{V}^{L} + 2 M
g_{V}^{L}|^{2} - \frac{1}{2} (3 +  \qom) |C_{A}^{L} +
m_{\mu}\qovm g_{A}^{L}|^{2}]\}.  \nonumber \eeq It can be seen
that if ${\bf {\hat P}_{\mu}} || {\bf \hat{q}}$,
$m_{\nu}\rightarrow 0$, one gives ${\bf <
S_{\nu}\cdot(\hat{P}_{\mu}\times \hat{q}) >}_{f} \equiv 0$ (Eq.
B1), and ${\bf <S_{\nu}\cdot \hat{ P}_{\mu}>}_{f} \; \mbox{(Eq.
B2) }\Rightarrow {\bf <S_{\nu}\cdot{\hat q}>}_{f}$ (Eq. D3).
\section{Nuclear observables and longitudinal neutrino polarization}
The formulas for  the longitudinal and  transverse components of
neutron polarization (T-even and T-odd components respectively),
and for the longitudinal neutrino  polarization (T-even quantity),
 in the case  of  non-vanishing neutrino mass $(m_{\nu}\not = 0)$,
 when the induced couplings are enclosed and ${\bf
{\hat P}_{\mu}}$, ${\bf \hat{q}}$ are  perpendicular to each
other, are as follows: \beq \label{jadr} {\bf <J_{n} \cdot
\hat{q}>}_{f} &=&
 \fip \{\mbox{} \qove |C_{A}^{L} + m_{\mu}\qovm  g_{A}^{L}|^{2}
- Re[(2\qom + \qove)(C_{V}^{L} + 2 M g_{V}^{L})(C_{A}^{L*} +
  m_{\mu}\qovm g_{A}^{L*})\\
   && \mbox{} - \me\qovm (C_{A}^{L} +
  m_{\mu}\qovm  g_{A}^{L})C_{S}^{R*}]\}, \nonumber
\\ {\bf <J_{n}\cdot({\hat P}_{\mu}\times{\hat q})>}_{f} & = &
\mbox{}\fip\pmo Im\{-(\qove + \qom)(C_{V}^{L} + 2 M
g_{V}^{L})(C_{A}^{L*} + m_{\mu}\qovm g_{A}^{L*}) \\ && \mbox{} -
\me\qovm (C_{V}^{L} + 2 M
  g_{V}^{L})C_{S}^{R*} \},\nonumber \\
{\bf <S_{\nu}\cdot{\hat q}>}_{f} & = & \fip\{-({3\over 2}\qove
+\qovm)|C_{A}^{L} + m_{\mu}\qovm  g_{A}^{L}|^{2} - ({1\over
2}\qove + \qovm)|C_{V}^{L} + 2 M g_{V}^{L}|^{2}
\\ && \mbox{} + {1\over 2}\qove \scr + Re[\qom(C_{V}^{L} + 2 M g_{V}^{L})(C_{A}^{L*} +
  m_{\mu}\qovm g_{A}^{L*})
  - \me\qovm(C_{V}^{L} + 2 M g_{V}^{L})C_{S}^{R*}]\}.\nonumber
 \eeq
 It can be seen that in these observables the occurrence of the
interference term between the standard $C_{V,A}^{L}$ couplings and
exotic $C_{S}^{R}$ coupling depends explicitly on the neutrino
mass. This dependence causes  the "conspiracy" of the interference
term and makes the measurement of the relative phase   between the
standard $C_{V,A}^{L}$ and exotic $C_{S}^{R}$ impossible because
$(m_{\nu}/E_{\nu})(q/2M)$ is very small.
\section{General form of nuclear observables and  of longitudinal neutrino polarization}
General results for the longitudinal and  transverse components
of neutron polarization and for the longitudinal neutrino
polarization, in the limit of vanishing neutrino mass $(m_{\nu}
\rightarrow 0)$, when the induced couplings are enclosed and
${\bf {\hat P}_{\mu}}$,  ${\bf \hat{q}}$ are not perpendicular to
each other, are as follows: \beq {\bf <J_{n} \cdot \hat{q}>}_{f}
& = &  \fip \{[-(1 + 2 \qom) Re((C_{V}^{L} + 2 M
g_{V}^{L})(C_{A}^{L*} +
  m_{\mu}\qovm g_{A}^{L*})) +  |C_{A}^{L} + m_{\mu}\qovm  g_{A}^{L}|^{2}] \nonumber\\
&  & \mbox{} + \pqp [Re((C_{V}^{L} + 2 M g_{V}^{L})(C_{A}^{L*} +
  m_{\mu}\qovm g_{A}^{L*}))  + |C_{A}^{L} + m_{\mu}\qovm  g_{A}^{L}|^{2} ]\}, \\
{\bf <J_{n}\cdot(\hat{P}_{\mu} \times \hat{q}) >}_{f} & = &
  \fip \pqkpj (1 + \qom) Im(
  (C_{V}^{L} + 2 M g_{V}^{L})(C_{A}^{L*} +
  m_{\mu}\qovm g_{A}^{L*})), \\
{\bf <S_{\nu}\cdot \hat{q}>}_{f} & = & \fip \{[\qom Re((C_{V}^{L}
+ 2 M g_{V}^{L})(C_{A}^{L*} + m_{\mu}\qovm
g_{A}^{L*})) + \frac{1}{2} \scr \\
&& \mbox{} - \frac{1}{2} (1 + \qom) |C_{V}^{L} + 2 M
g_{V}^{L}|^{2}  - \frac{1}{2} (3  +   \qom) |C_{A}^{L} +
m_{\mu}\qovm  g_{A}^{L}|^{2} ] \nonumber \\ & & \mbox{} + \pqp [
\qom Re((C_{V}^{L} + 2 M g_{V}^{L})(C_{A}^{L*} +
 m_{\mu}\qovm g_{A}^{L*})) \nonumber \\
 & & \mbox{} + \frac{1}{2}(\scr
 + (1 + \qom) |C_{V}^{L} + 2 M g_{V}^{L}|^{2} +
 (\qom - 1) |C_{A}^{L} + m_{\mu}\qovm  g_{A}^{L}|^{2})]\}.\nonumber \eeq
 \section{Four-vector neutrino polarization and Michel-Wightman density matrix}
 The formulas for the 4-vector initial neutrino polarization in its rest
 frame and for the initial neutrino moving  with the momentum ${\bf q}$,
 respectively, are as follows:
\beq S & = & (0,\mbox{\boldmath $\eta_{\nu}$}),\\
 S' & = & \frac{\mbox{\boldmath $\eta_{\nu}$}\cdot{\bf q}}{E_{\nu}}\cdot
\frac{1}{m_{\nu}} \left(
\begin{array}{c}  E_{\nu}\\ {\bf q} \end{array} \right)
 + \left(
\begin{array}{c}  0\\ \mbox{\boldmath $\eta_{\nu}$}  \end{array} \right) -
\frac{\mbox{\boldmath $\eta_{\nu}$}\cdot{\bf
q}}{E_{\nu}(E_{\nu}+m_{\nu})} \left( \begin{array}{c}  0\\ {\bf q}
\end{array} \right),
 \eeq
 where $\mbox{\boldmath $\eta_{\nu}$}$ - the full 3-vector of the
 initial neutrino polarization in its rest frame.
 The formulas for the  Michel-Wightman density matrix in the
 case of the polarized neutrinos with the non-zero neutrino mass and in the limit of vanishing
 neutrino mass, respectively,  are as follows:
 \beq
\Lambda_{\nu}^{(s)}=\sum_{\nu}u_{\nu}\overline{u}_{\nu} \sim
[1+\gamma_{5}(S^{'\mu}\gamma_{\mu})][(q^{\mu}\gamma_{\mu}) +
m_{\nu}] & = & \mbox{} [(q^{\mu}\gamma_{\mu}) + m_{\nu} +
\gamma_{5}(S^{'\mu}\gamma_{\mu})(q^{\mu}\gamma_{\mu}) +
\gamma_{5}(S^{'\mu}\gamma_{\mu}) m_{\nu}],  \\
(S^{'\mu}\gamma_{\mu}) &=& \frac{\mbox{\boldmath
$\eta_{\nu}$}\cdot{\bf q}}{E_{\nu}m_{\nu}}(q^{\mu} \gamma_{\mu}) -
(\mbox{\boldmath $\eta_{\nu}$} - \frac{(\mbox{\boldmath
$\eta_{\nu}$}\cdot{\bf q}){\bf q}}{E_{\nu}(E_{\nu}+
m_{\nu})})\cdot\mbox{\boldmath$\gamma$}, \\
(S^{'\mu}\gamma_{\mu})(q^{\mu}\gamma_{\mu}) &=&
\frac{m_{\nu}}{E_{\nu}}\mbox{\boldmath $\eta_{\nu}$}\cdot{\bf q} -
(\mbox{\boldmath $\eta_{\nu}$} - \frac{(\mbox{\boldmath
$\eta_{\nu}$}\cdot{\bf q}){\bf q}}{E_{\nu}(E_{\nu}+
m_{\nu})})\cdot \mbox{\boldmath $\gamma$}(q^{\mu}\gamma_{\mu}),
\\ (S^{'\mu}\gamma_{\mu}) m_{\nu} &=& \frac{\mbox{\boldmath
$\eta_{\nu}$}\cdot{\bf q}}{E_{\nu}}(q^{\mu}\gamma_{\mu}) -
m_{\nu}(\mbox{\boldmath $\eta_{\nu}$} - \frac{(\mbox{\boldmath
$\eta_{\nu}$}\cdot{\bf q}){\bf q}}{E_{\nu}(E_{\nu}+
m_{\nu})})\cdot \mbox{\boldmath $\gamma$}, \\
\lim_{m_{\nu}\rightarrow 0}
[1+\gamma_{5}(S^{'\mu}\gamma_{\mu})][(q^{\mu}\gamma_{\mu}) +
m_{\nu}] & = & \mbox{} [1 + \gamma_{5}[\frac{\mbox{\boldmath
$\eta_{\nu}$} \cdot{\bf q}}{|{\bf q}|} - (\mbox{\boldmath
$\eta_{\nu}$} - \frac{(\mbox{\boldmath $\eta_{\nu}$}\cdot{\bf
q}){\bf q}}{|{\bf q}|^{2}})\cdot \mbox{\boldmath
$\gamma$}]](q^{\mu}\gamma_{\mu}).  \eeq
\end{widetext}
%
\begin{acknowledgments}
I am greatly indebted to  Prof.\  S.\ Ciechanowicz for many
useful and helpful discussions. This work was supported in part
by the Foundation for  Polish Science.
\end{acknowledgments}

\end{document}